\documentclass{article}
\usepackage{slashed,feynmf,epsfig,amsmath,amssymb,enumitem}
\usepackage[papersize={8.5in,11in}]{geometry}
\begin{document}
\title{Unitarity In An Alternative Electroweak Theory}
\author{J. W. Moffat\\~\\
Perimeter Institute for Theoretical Physics, Waterloo, Ontario N2L 2Y5, Canada\\
and\\
Department of Physics and Astronomy, University of Waterloo, Waterloo,\\
Ontario N2L 3G1, Canada}
\maketitle
\begin{abstract}
An electroweak (EW) model has been investigated~\cite{Moffat} in which the energy $E < \mu=\sqrt{\lambda}M_W$, where $\lambda$ is a gauge parameter and $M_W$ is the $W$ boson mass. For large enough $\lambda$ the scalar boson mass $\mu=\sqrt{\lambda}M_W$ can be heavy enough to avoid detection in LHC experiments. The theory is perturbatively renormalizable for the decoupled scalar interactions. The Born approximation tree graph unitarity can be ensured by postulating that the effective coupling constant $g_{\rm eff}(s)$ vanishes as $1/\sqrt{s}$ or faster as $s\rightarrow\infty$, predicting that the EW interactions become weaker at high energies, and longitudinally polarized $W_L W_L\rightarrow W_L W_L$ scattering does not violate Born approximation tree graph unitarity.
\end{abstract}

\begin{fmffile}{ewunifigs}

\section{Introduction}

The ATLAS and CMS collaborations have released the preliminary results of their search for the Standard Model Higgs boson at LHC based on almost 5 $fb^{-1}$ of data~\cite{CERN}. The results in the search for a Higgs particle remain inconclusive. However, the range of Higgs particle mass has been excluded from $M_H\sim 130$ GeV to $M_H\sim 600$ GeV, leaving a $15$ GeV window between the LEP2 upper exclusion bound $M_H\sim 115$ GeV and $M_H\sim 130$ GeV of possible detection of the Higgs particle. If the Higgs particle is not detected, then we must consider revising at a fundamental level the electroweak (EW) model of Weinberg and Salam~\cite{Weinberg,Salam,Aitchison}. This may require a revision of our ideas about QFT. A previously published EW theory without a Higgs particle and a quantum gravity theory~\cite{Moffat2,Moffat3,Moffat4,Moffat5} were based on nonlocal interactions and the EW theory led to finite amplitudes and cross sections that can be tested at the LHC.

In a recent paper~\cite{Moffat}, an effective EW model was based on a gauge invariant action with {\it local interactions} using a Stueckelberg formalism~\cite{Stueckelberg,Ruegg}. The gauge invariance of the Lagrangian leads to a renormalizable EW theory, provided the scalar fields in the Lagrangian decouple at high energies rendering a scalar spin-0 boson undetectable at present accelerator energies. The model contains only the observed particles, namely, 12 quarks and leptons, the charged $W$ boson, the neutral $Z$ boson and the massless photon and gluon without the Higgs particle. In the following, we investigate how to maintain Born approximation unitarity for the tree graph calculation of $W_L+W_L\rightarrow W_L+W_L$ longitudinally polarized scattering above an energy of 1-2 TeV. To guarantee that unitarity is not violated in the effective energy range $E < \mu=\sqrt{\lambda}M_W$, we postulate that the effective coupling $g_{\rm eff}(s)$ decreases as $\sim 1/\sqrt{s}$ or faster as $\sqrt{s}\rightarrow\infty$, where $\sqrt{s}$ is the center-of-mass energy.

In our EW theory, there is no attempt to explain the origin of elementary particle masses. This is particularly true for the $W$ and $Z^0$ bosons. The gauge invariant Lagrangian is given by
\begin{align}
\label{GaugeEWLagrangian}
{\cal L}_{\rm EW}=\sum_f {\bar f}i\slashed{D}^W f+\sum_f{\bar f}i\slashed{D}^Z f
+\sum_f{\bar f}i\slashed{D}^A f-\frac{1}{2}W^{+\mu\nu}W^{-}_{\mu\nu}-\frac{1}{4}Z_{\mu\nu}Z^{\mu\nu}-\frac{1}{4}F^{\mu\nu}F_{\mu\nu}\nonumber\\
+\frac{1}{2}(M_ZZ_\mu-\partial_\mu\beta)(M_Z Z^\mu-\partial^\mu\beta)
+(M_WW^+_\mu-P\partial_\mu\sigma)(M_WW^{-\mu}-P\partial^{\mu}\sigma)+{\cal L}_{m_f},
\end{align}
where $\beta$ and $\sigma$ are scalar gauge fields and $P$ is a function of $\sigma$. The Lagrangian (\ref{GaugeEWLagrangian}) is invariant under the infinitesimal gauge transformations
\begin{equation}
Z_\mu\rightarrow Z_\mu+\partial_\mu\nu,\quad \beta\rightarrow \beta+M_Z\nu,
\end{equation}
and
\begin{equation}
W_\mu\rightarrow W_\mu+{\cal D}^W_\mu\chi,\quad \sigma\rightarrow \sigma+Q\chi.
\end{equation}
Here, $Q$ is an unknown function of $\chi$, ${\cal D}^W_\mu$ is a covariant differential operator and $\slashed{D}^{W,Z,A}=\gamma^\mu D^{W,Z,A}_\mu$. We have $W^+_\mu=(W^-_\mu)^\dagger$ where
\begin{equation}
W^+_\mu=\frac{1}{\sqrt{2}}(W^1_\mu-iW^2_\mu),\quad W^-_\mu=\frac{1}{\sqrt{2}}(W^1_\mu+iW^2_\mu),
\end{equation}
and
\begin{equation}
\label{WZequation}
W^+_{\mu\nu}=D^W_\mu W^+_\nu-D^W_\nu W^+_\mu,\quad Z_{\mu\nu}=\partial_\mu Z_\nu -\partial_\nu Z_\mu,
\end{equation}
where $W^+_{\mu\nu}=(W^-_{\mu\nu})^\dagger$. The non-Abelian field $W^a_{\mu\nu}$ is given by
\begin{equation}
W^a_{\mu\nu}=\partial_\mu W^a_\nu-\partial_\nu W^a_\mu+gf^{abc}W^b_\mu W^c_\nu.
\end{equation}

The fermion mass Lagrangian is
\begin{equation}
\label{fermionmass}
{\cal L}_{m_f}=-\sum_{\psi_L^i,\psi_R^j}m_{ij}^f(\bar\psi_L^i\psi_R^j + \bar\psi_R^i\psi_L^j),
\end{equation}
where $\psi_{L,R}=P_{L,R}\psi$, $P_{L,R}=\frac{1}{2}(1\mp\gamma_5)$ and $m_{ij}^f$ denotes the fermion masses. Eq. (\ref{fermionmass}) can incorporate massive neutrinos and their flavor oscillations.

We have not included in the Lagrangian $(\ref{GaugeEWLagrangian})$ the standard scalar field Higgs contribution:
\begin{equation}
\label{scalarLagrangian}
{\cal L}_\phi=\large\vert(i\partial_\mu - gT^aW^a_\mu - g'\frac{Y}{2}B_\mu)\phi\large\vert^2-V(\phi),
\end{equation}
where $\vert...\vert^2=(...)^\dagger(...)$. Moreover,
\begin{equation}
\label{potential}
V(\phi)=\mu_H^2\phi^\dagger\phi +\lambda_H(\phi^\dagger\phi)^2,
\end{equation}
and $B_\mu$ is the neutral vector boson that couples to weak hypercharge, $\mu^2_H < 0$ and $\lambda_H > 0$. The $W$ and $Z$ masses in Eq. (\ref{GaugeEWLagrangian}) are the experimental masses. We do not begin with a massless Lagrangian and then break the $SU(2)$ symmetry through a spontaneous symmetry breaking of the vacuum.

The Stueckelberg gauge invariance is of the ``hidden'' symmetry type similar to the gauge invariance of the standard model Lagrangian after spontaneous symmetry breaking of the vacuum. Gauge invariance is a necessary requirement to guarantee that the theory is renormalizable. Our effective EW theory is not ultraviolet (UV) complete. We have to demand that we apply the theory in the restricted energy region $E < \mu=\sqrt{\lambda} M_W$ in which there are no scalar bosons in either the initial or final state of the S matrix. Thus, the non-renormalizable scalar field interactions are decoupled from the non-Abelian $W^{\pm}$ interactions. This decoupling of the scalar field interactions follows automatically for the neutral $Z^0$ boson Abelian interactions and this $U(1)$ sector of the theory is renormalizable and unitary. For a sufficiently large $\lambda$ parameter the scalar boson has a mass $\mu=\sqrt{\lambda}M_W$ that makes is heavy enough to avoid detection by LHC experiments. In the restricted effective energy range the S matrix is unitary:
\begin{equation}
S^\dagger=S^{-1}.
\end{equation}
However, this does not in itself guarantee the Born tree graph approximation is unitary. It was first shown by Llewellyn Smith~\cite{Smith} that for a single elementary Higgs boson only the Weinberg-Salam theory~\cite{Weinberg,Salam}, based on the spontaneously broken $SU(2)\times U(1)$ guarantees that the tree graph unitary for longitudinally polarized $f{\bar f}\rightarrow W_L^+ W_L^-$ and $W_L^+ + W_L^-\rightarrow W_L^+ + W_L^-$ is satisfied. This was also demonstrated by Cornwall, Levin and Tiktopoulos~\cite{Cornwall}.

The Feynman tree graphs for the $W_LW_L$ scattering are:
\vskip 16pt\noindent
\begin{equation}
\parbox{1in}{\begin{fmfgraph*}(50,30)
\fmfleftn{i}{2}\fmfrightn{o}{2}
\fmf{boson}{i1,v1}
\fmf{boson}{i2,v1}
\fmf{boson,label=$\gamma/Z^0$,lab.side=left}{v1,v2}
\fmf{boson}{v2,o1}
\fmf{boson}{v2,o2}
\fmflabel{$W^{+}$}{i1}
\fmflabel{$W^{-}$}{i2}
\fmflabel{$W^{+}$}{o1}
\fmflabel{$W^{-}$}{o2}
\end{fmfgraph*}}
+~~~~~~~~~~
\parbox{1in}{\begin{fmfgraph*}(40,40)
\fmfbottomn{i}{2}\fmftopn{o}{2}
\fmf{boson}{i1,v1}
\fmf{boson}{i2,v1}
\fmf{boson,label=$\gamma/Z^0$}{v1,v2}
\fmf{boson}{v2,o1}
\fmf{boson}{v2,o2}
\fmflabel{$W^{+}$}{i1}
\fmflabel{$W^{+}$}{i2}
\fmflabel{$W^{-}$}{o1}
\fmflabel{$W^{-}$}{o2}
\end{fmfgraph*}}\label{eq:WWST}
\end{equation}
\vskip 8pt\noindent
We add to these graphs the $W$ boson contact interaction graph:
\vskip 16pt\noindent
\begin{align}
\parbox{0.55in}{\begin{fmfgraph*}(40,30)
\fmfleftn{i}{2}\fmfrightn{o}{2}
\fmf{boson}{i1,v1}
\fmf{boson}{i2,v1}
\fmf{boson}{v1,o1}
\fmf{boson}{v1,o2}
\fmflabel{$W^-_\lambda$}{i1}
\fmflabel{$W^+_\mu$}{i2}
\fmflabel{$W^-_\rho$}{o1}
\fmflabel{$W^+_\nu$}{o2}
\end{fmfgraph*}}\label{eq:4W}\\
&\nonumber
\end{align}
\vskip 6pt\noindent

The amplitude for the longitudinally polarized $W$ scattering is given by~\cite{MoffToth}:
\begin{equation}
\label{WWamplitude}
{\cal A}(W_LW_L\rightarrow W_LW_L)=g^2\left[\frac{\cos\theta+1}{8M_W^2}s+{\cal O}(1)\right],
\end{equation}
where $g$ is the low energy weak coupling constant. Eq. (\ref{WWamplitude}) clearly violates unitarity for large $s$. However, this behavior is corrected in the standard Weinberg-Salam model by the addition of the $s$-channel Higgs process:
\vskip 16pt\noindent
\begin{align}
\parbox{0.55in}{\begin{fmfgraph*}(50,30)
\fmfleftn{i}{2}\fmfrightn{o}{2}
\fmf{boson}{i1,v1}
\fmf{boson}{i2,v1}
\fmf{dashes,label=$H$}{v1,v2}
\fmf{boson}{v2,o1}
\fmf{boson}{v2,o2}
\fmflabel{$W^-_\lambda$}{i1}
\fmflabel{$W^+_\mu$}{i2}
\fmflabel{$W^-_\rho$}{o1}
\fmflabel{$W^+_\nu$}{o2}
\end{fmfgraph*}}\label{eq:WWH}\\
&\nonumber
\end{align}
\vskip 8pt\noindent
In the high energy limit we get
\begin{equation}
\label{Higgscontribution}
{\cal A}_H=-g^2\left[\frac{\cos\theta+1}{8M_W^2}s+{\cal O}(1)\right],
\end{equation}
which cancels the bad high energy behavior in (\ref{WWamplitude}). The resulting scattering amplitude in the standard EW model based on EW spontaneous symmetry breaking, including the Higgs particle exchange graph, is given by
\begin{equation}
\label{WWstandardmodel}
{\cal A}_\mathrm{SM}(W_LW_L\rightarrow W_LW_L)=g^2\left[\frac{\cos^2\theta+3}{4\cos\theta_w^2(1-\cos\theta)}-\frac{M_H^2}{2M_W^2}+{\cal O}(s^{-1})\right].
\end{equation}
We observe that if we integrate over the scattering angle $\theta$ in (\ref{WWstandardmodel}) to obtain the cross section, then there is no energy dependence to order ${\cal O}(s^{-1})$. The fundamental Higgs mass cannot be bigger than $M_H\sim 600$ GeV to prevent a break down of perturbation theory and unitarity.

In our EW model the standard Higgs mechanism generated by spontaneous symmetry breaking of the vacuum is absent, and we do not have a Higgs boson contribution corresponding to (\ref{Higgscontribution}). Moreover, the scalar bosons do not contribute to the S matrix in our restricted energy range $E < \mu=\lambda^{1/2}M_W$, so we solve the unitarity problem by asserting that we have instead of (\ref{WWamplitude}):
\begin{equation}
\label{WWamplitude2}
{\cal A}(W_LW_L\rightarrow W_LW_L)=g^2_{\rm eff}(s)\left[\frac{\cos\theta+1}{8m_W^2}s+{\cal O}(1)\right],
\end{equation}
where
\begin{equation}
g_{\rm eff}(s)=g\biggl[1+(\sqrt{s}/{\cal M})^n\biggr]^{-1}.
\end{equation}
Here, ${\cal M}$ is a parameter that determines the energy scale of the running of the effective $g_{\rm eff}(s)$ and $n\geq 1$. Now $g_{\rm eff}$ decreases fast enough as $\sqrt{s}\rightarrow\infty$, so that the UV unitarity violation can be avoided for the tree graph approximation and the scattering amplitude and the cross section can remain below the unitarity bound for ${\cal M} > 1-2$ TeV. If the running of $g_{\rm eff}(s)$ decreases fast enough as $\sqrt{s}\rightarrow\infty$, then the EW scattering amplitudes and cross sections can become weaker with increasing energy, implying that the vector bosons $W$ and $Z$ satisfy a form of asymptotic safety.

Calmet~\cite{Calmet} has proposed that in an alternative EW model the EW scale $v_{\rm eff}$ runs with energy. By substituting into (\ref{WWamplitude}) the result $M_W=(1/2)gv$, derived from the standard spontaneous symmetry breaking of $SU(2)\times U(1)$, where $v$ is the EW energy scale $v\sim 246$ GeV, we get
\begin{equation}
v_{\rm eff}=v\biggl(1+\frac{\omega}{8\pi}\frac{M^2}{v^2}\biggr)^{1/2},
\end{equation}
where $M$ is a running mass scale, $\omega$ is a non-perturbative parameter and for spontaneous symmetry breaking $v=\langle\phi_H\rangle_0$ where $\phi_H$ is the scalar Higgs field. In this model, the running of $v_{\rm eff}$ can prevent a violation of tree graph unitarity. In both our EW model and in the Calmet model, the weak interactions can become weaker with increasing $\sqrt{s}$ corresponding to asymptotic safety.

\section{Conclusions}

In our model without the scalar boson interactions in the restricted energy range $E < \mu=\lambda^{1/2}M_W$ for which the scalar boson mass $\mu$ is sufficiently heavy to avoid detection by the LHC, the theory will be both renormalizable and unitary. In composite models of weak interactions such as technicolor~\cite{Technicolor} and the postulate of heavy resonances at energies above 1 TeV~\cite{Grojean}, the weak interactions are predicted to become strong, positing a unitarization of the strongly bound state particles. Experiments conducted at the LHC can test the predictions of our EW model and alternative models by measuring the weak interaction cross sections and determining whether they in fact do grow weaker than predicted in the standard Weinberg-Salam model, or whether they grow stronger as in composite models.

\end{fmffile}

\section*{Acknowledgements}

I thank Viktor Toth for helpful and stimulating discussions. This research was supported by the John Templeton Foundation.
Research at the Perimeter Institute for Theoretical Physics is supported by the Government of Canada through NSERC and by the Province of Ontario through the Ministry of Research and Innovation (MRI).

\end{document}